\documentclass[amsmath,twocolumn,showpacs]{revtex4}
\usepackage{epsfig}
\usepackage{amsfonts}

\begin{document}

\newcommand{\beq}{\begin{equation}}
\newcommand{\eeq}{\end{equation}}
\newcommand{\beqa}{\begin{eqnarray}}
\newcommand{\eeqa}{\end{eqnarray}}
\newcommand{\bra}[1]{\ensuremath{{\langle{#1}|}}}
\newcommand{\ket}[1]{\ensuremath{{|{#1}\rangle}}}
\newcommand{\braket}[2]{\ensuremath{{\langle{#1}|{#2}\rangle}}}
\newcommand{\bbra}[1]{\ensuremath{{\langle\!\langle{#1}|}}}
\newcommand{\kket}[1]{\ensuremath{{|{#1}\rangle\!\rangle}}}
\newcommand{\bbrakket}[2]{\ensuremath{{\langle\!\langle{#1}|{#2}\rangle\!\rangle}}}
\newcommand{\average}[1]{\ensuremath{{\langle{#1}\rangle}}}
\newcommand{\Tr}[1]{{\rm Tr}\left({#1}\right)}

\title{Local control theory for unitary transformations: Application to quantum computing without leakage}

\author{Shlomo E. Sklarz}
\author{David J. Tannor}
\affiliation{Department of Chemical Physics, Weizmann Institute of Science \\
76100, Rehovot, Israel.\\ Tel 972-8-9343723, Fax 972-8-9344123}

\begin{abstract}
We present a local optimal control strategy to produce desired unitary transformations. Unitary transformations are central to all quantum computational algorithms. Many realizations of quantum computation use a submanifold of states, comprising the quantum register, coupled by an external driving field to a collection of additional mediating excited states.
Previous attempts to apply control theory to induce unitary transformations on the quantum register, while successful, produced pulses that drive the population out of the computational register at intermediate times.
Leakage of population from the register is undesirable since often the  states outside the register are prone to decay and decoherence, and populating them causes a decrease in the final fidelity. 
In this work we devise a local optimal control method for achieving target unitary transformations on a quantum register, while avoiding intermediate leakage out of the computational submanifold.  The technique exploits a phase locking of the field to the system such as to eliminate the undesirable excitation.
This method is then applied to produce an $SU(6)$ Fourier transform on the vibrational levels of the ground electronic  state of the Na$_2$ molecule. The emerging mechanism uses two photon resonances to create a transformation on the quantum register while blocking one photon resonances to excited states.
\end{abstract}
\maketitle
%%%%%%%%%%%%%%%%%%%%%%%%%%%%%%
%\section{Introduction}
 
In recent years there has been growing interest in the possibility of realizing quantum computers.   Any such implementation must consist of a quantum register comprised of a selection of quantum states on which the computational operations can be performed. A physical realization of a quantum computer should therefore be able to produce these unitary transformations on the register by the use of external driving fields \cite{DiV95}. 

The general paradigm of quantum computing is to break up every computational operation into a sequence of simple unitary operations called quantum gates which can be considered the basic building blocks of quantum computation \cite{Nielsen00}.  
As the register size grows the number of quantum gates required to construct an arbitrary unitary operation  increases rapidly, and therefore the total fidelity, which depends on the accumulation of errors at each step, decreases drastically \cite{Lioyd94}. 
It has been shown  in \cite{Rangan01,Tesch02,Palao02} that Optimal Control theory (OCT) methods \cite{Rice00,Shapiro03,Peirce88,Kosloff89,Ohtsuki03} can be utilized in order to calculate a  field that will directly induce an arbitrary target transformation. 
The OCT approach  eliminates the need to decompose the unitary operator into fundamental operations; however the emerging fields are complicated and  drive the population out of the computational submanifold at intermediate times. 
This is undesirable since in physical realizations the excited states are prone to decay and decoherence, and therefore their population at intermediate times causes a decrease in the final fidelity. 

In this paper we use a variant of control theory, which we refer to as ``local control'', to achieve a target unitary transformation on a quantum register with no intermediate leakage out of the quantum register. 
The technique exploits a phase locking of the field to the system such as to eliminate undesirable excitations.
We demonstrate this approach by obtaining fields
to produce an $SU(6)$ FT on a  quantum register consisting of a submanifold of vibrational levels on the ground electronic state of a diatomic  molecule \cite{Zadoyan01,Amitay02} without populating the excited electronic vibrational states.
The emerging mechanism uses two photon resonances to transform  the quantum register while blocking one photon resonances to excited states.
To some extent the method can be viewed as a systematic generalization of the methods of Monroe {\it et al} \cite{Monroe95} and S{\o}rensen and M{\o}lmer \cite{Sorensen99} to an arbitrary number of qubits and arbitrary unitary transformations. The more complicated fields presented here can be realized using optical pulse shaping techniques \cite{Weiner00,Brixner03.2,Oron03}.

%%%%%%%%%%%%%%%%%%%%%%%%%%%%%%%%%%%%%%%%%%%%%%
%\section{Definition of the unitary control problem}
The model Hamiltonian consists of a free part $\hat H_0$ and an interaction part $\hat V=-\hat \mu E(t)-\hat \mu^\dagger E^*(t)$ controlled by an external field $E(t)$ through the dipole operator $\hat \mu=\sum_{ij}\mu_{ij}\ket{i}\bra{j}$, with  $\mu_{ij}$ the coupling strengths between the ground and excited states \ket{i} and \ket{j} respectively, 
\beq
\hat H=\hat H_0-\hat \mu E(t)-\hat \mu^\dagger E^*(t).
\eeq
The system evolution can be described by a time dependent unitary transformation $U(t)$, the dynamics of which is governed by the Schr\"odinger equation,
\beq
\frac{\partial U(t)}{\partial t}= -{i \over \hbar}\hat HU(t), 
\eeq
with the initial condition $U(0)=\Bbb{I}$. In order to eliminate the free Hamiltonian motion it is common practice to switch to  the interaction picture Hamiltonian
\beqa
\hat H\to\tilde H&=& \exp\{\frac{i}{\hbar}H_0t\}\hat V\exp\{-\frac{i}{\hbar}H_0t\}\nonumber\\
&=&-\tilde\mu(t)E(t)-\hat \mu^\dagger(t) E^*(t). 
\eeqa
with $\tilde\mu(t)\equiv\exp\{\frac{i}{\hbar}H_0t\}\hat\mu\exp\{-\frac{i}{\hbar}H_0t\}$.

The desired computation  is represented by a certain unitary transformation $O_r$, which is defined to operate solely on the restricted subspace constituting the quantum register and which  must be obtained by the system at the final time $T$. 
Denoting $P_r$ to be a projection onto the restricted subspace we define $U_r(t)=P_r U(t) P_r$ to be the portion of the evolution operator  acting on the restricted computational subspace. The goal is therefore to obtain a control field $E(t)$ which will maximize the overlap, $J=|{\rm Tr}(O_r^\dagger U_r(T))|^2$, between the target $O_r$ and $U_r(t)$ at the final time, subject to the constraint that the computational manifold population $C={\rm Tr}(U_r^\dagger(t)U_r(t))$ remains fixed throughout. This constraint enforces the elimination of leakage from the computational manifold throughout the evolution. 

%\section{The Local control Method of optimization}
We introduce, here, a local control method which continuously increases the objective $J$ while simultaneously holding the constraint $C$ fixed. At each time step, the algorithm 
constructs a field  $E(t)$ that fulfills the two conditions. 
The constraint  $dC/dt=0$, determines the direction of the field vector $E(t)$ in the complex plane. 
The sign of the field, however, remains free and is chosen such as to make the time derivative of $J$ positive, i.e.  $dJ/dt\ge 0$, ensuring an increase in the objective at the next time step.

We now derive the equations determining the direction and magnitude of $E(t)$ at each step. Note first that since $U_r\equiv P_r U P_r$ we have 
$$\dot U_r=P_r\frac{1}{i\hbar}\tilde HUP_r.$$ 
The equation for the constraint is 
\beqa\label{eq:Ca}
\dot C&=&\frac{d}{dt}{\rm Tr}(U^\dagger_r U_r)\nonumber\\
&=&2{\rm Re}\left\{{\rm Tr}(U^\dagger_rP_r\frac{1}{i\hbar}\tilde HUP_r)\right\}\nonumber\\
&=&-\frac{2}{\hbar}{\rm Im}\{{\rm Tr}(U^\dagger_rP_r\tilde \mu(t)UP_r)E(t)\nonumber\\
&&+{\rm Tr}(U^\dagger_rP_r\tilde \mu^\dagger(t)UP_r)E^*(t)\}=0
\eeqa
Noticing that the trace operation is invariant under transposition of its argument and that the imaginary part changes sign under complex conjugation we can take the adjoint of the second term and write
\beqa\label{eq:Cb}
\dot C&=&-\frac{2}{\hbar}{\rm Im}\{{\rm Tr}(U^\dagger_rP_r\tilde \mu(t)UP_r)E(t)\nonumber\\
&&-{\rm Tr}(P_r U^\dagger\tilde \mu(t)P_rU_r)E(t)\}\nonumber\\
&=&-2{\rm Im}\left\{gE(t)\right\}=0
\eeqa
where $g=Tr(U^\dagger_rP_r\tilde \mu(t)UP_r - P_r U^\dagger\tilde \mu(t)P_rU_r)$. One can now enforce fulfillment of the constraint, eq. (\ref{eq:Cb}), by choosing the electric field to be in the direction of $g^*$ with (real) proportionality constant $\alpha$ namely
\beq
E(t)=\alpha \epsilon_g^*,
\eeq
with $\epsilon^*_g\equiv g^*/|g|$ denoting the direction of $g^*$ in the complex plane. 
\begin{figure}[ht]
\epsfig{file=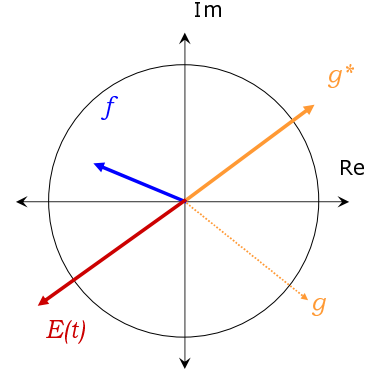,clip=,width=2in}
\caption[comp_fg]{\label{comp_fg} The quantities $g^*$ and $(f\cdot g)$ determine respectively the direction and magnitude of $E(t)$ in the complex plane. }
\end{figure}
We now use the freedom in choosing the sign of $\alpha$ to assure monotonic increase in the objective.
\beqa
\dot J&=&\frac{d}{dt}|{\rm Tr}(O^\dagger_rU_r)|^2\nonumber\\
&=&2{\rm Re}\left\{{\rm Tr}(O^\dagger_rU_r)^*\frac{d}{dt}{\rm Tr}(O^\dagger_rU_r)\right\}\nonumber\\
&=&2{\rm Re}\left\{\eta^*{\rm Tr}(O^\dagger_rP_r\frac{1}{i\hbar}\tilde HUP_r)\right\}\nonumber\\
&=&\alpha 2{\rm Re}\left\{f\epsilon_g^*\right\}\ge 0,
\eeqa   
where we define $\eta\equiv Tr(O^\dagger_rU_r)$ and $f=-\frac{1}{i\hbar}{\rm Tr}\left\{\eta^*O^\dagger P_r\hat\mu(t)UP_r-\eta P_r U^\dagger\hat\mu(t)P_rO_r\right\}$. Note that in advancing to the last line we have employed the same  sequence of steps as in eq. (\ref{eq:Ca}) and (\ref{eq:Cb}). In order to enforce an increase in $J$, we choose
\beqa
\alpha&=&Re\left\{fg^*\right\}\nonumber\\
&\equiv&(f\cdot g),
\eeqa
with  $(f\cdot g)$ denoting a scalar product in the complex plane, thus guaranteeing that $dJ/dt\ge0$.  The electric field is therefore chosen as 
\beq
E(t)=(f\cdot g)\epsilon^*_g,
\eeq
It is important that $E(t)$ depend on the magnitude of $g^*$ and not only on its  direction $\epsilon^*_{g}$, since for a vanishing $|g|$ the direction $\epsilon^*_g$ becomes undefined and numerically unstable. We therefore wish that $E$ be proportional to $|g|$ such that a phase jump in $g$  be accompanied by a vanishing of $E$, thus avoiding abrupt phase jumps in the emerging field. 

Note that there is still freedom in determining the magnitude of $E(t)$, which can be utilized to control the order of magnitude of the  field strength by multiplying by a positive envelope function, $S(t)$. It is sometimes necessary for numerical reasons to limit the maximum allowed field such that $|E(t)|<E_{\max}$. This can be achieved by transforming $\alpha\to E_{\max}\tanh(\alpha/E_{\max})$ which preserves the sign of $\alpha$ but saturates at constant value, $E_{\max}$.    

Summing up, our algorithm requires   calculating  $g$ and $f$ at each time step and choosing a field according to
\beq
E(t)=E_{\max}\tanh\left(S(t)(f\cdot g)/E_{\max}\right)\epsilon^*_g.
\eeq
In order to begin the control process it is necessary to seed a small fraction of population into the excited states at initial time. This can be done  by exciting the system with a weak pulse tuned to the optical transition. Alternatively, one can begin with an initial condition slightly rotated from the identity. As the seeding is negligibly small, the emerging field will produce the target unitary transformation also when applied to the `pure' initial condition $U(0)=\Bbb{I}$ as required. 
\begin{figure}[ht]
\epsfig{file=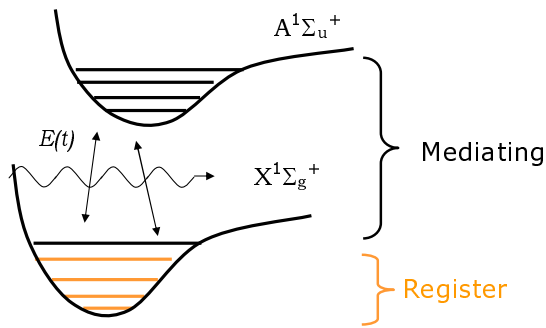,clip=,width=3in}
\caption[Na2]{\label{Na2} Schematic picture of Na$_2$ molecular model consisting of ground ($X^1\Sigma^+_g$) and excited ($A^1\Sigma^+_u$) electronic potential energy curves. A submanifold of the ground state levels constitutes  the quantum register. The remaining levels serve as  mediators which couple the register levels through interaction with the electric field. The goal is to produce a unitary transformation on the register levels with no intermediate population of the mediating states.}
\end{figure}

\begin{figure}[ht]
\epsfig{file=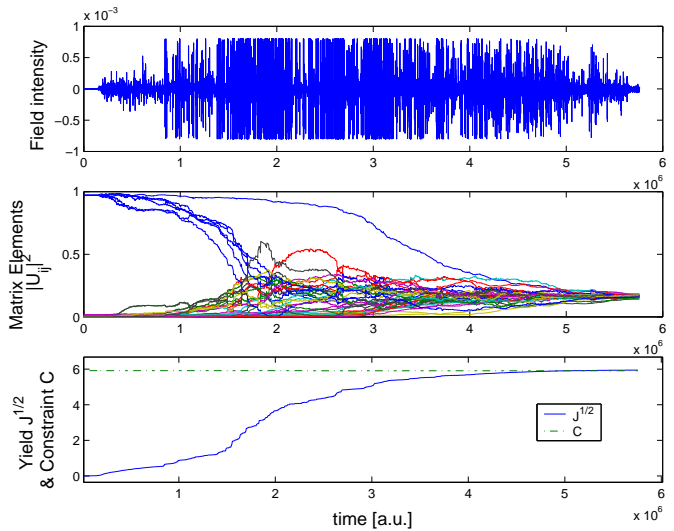,clip=,width=3.5in}
\caption[Na2_su6_6]{\label{Na2_su6_6}Top: The control field obtained by the algorithm. Middle: Controlled evolution of the elements of the unitary propagator (absolute values). Bottom: Evolution of objective $\sqrt{J}$ (solid) and constraint $C$ (dashed) showing monotonic increase in the former while keeping the latter constant.}
\end{figure}

\begin{figure*}[bht]
\epsfig{file=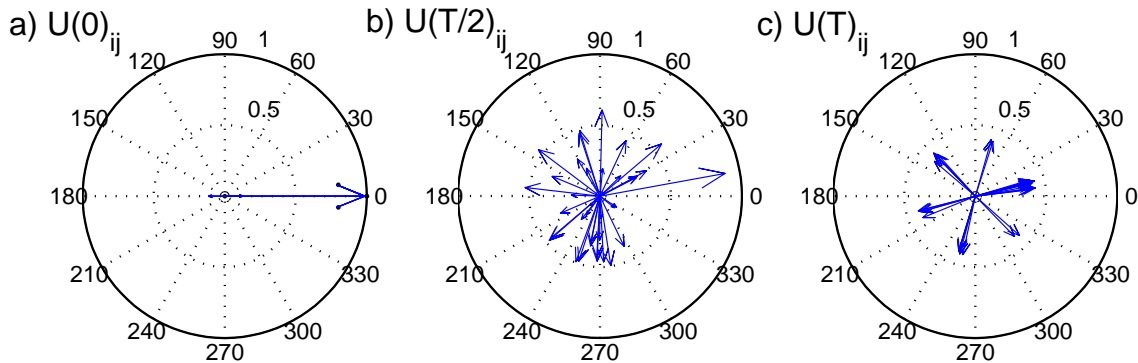,clip=,height=2in}
\caption[Na2_su6_6_compas]{\label{Na2_su6_6_compas}The elements of the unitary propagator at times a) $t=0$, b) $t=T/2$ and c) $t=T$ as vectors in the complex plane.}
\end{figure*}

%%%%%%%%%%%%%%%%%%%%%%%%%%%%%%%%%
%\section{Application to the Creating of an SU(6) Fourier transform in an N\lowercase{a}$_2$ molecular model} 
We apply the local control technique described above to the implementation of an $SU(N_r)$ Fourier transform on the ground potential surface of a two-electronic surface model of an Na$_2$ molecule. The first $N_r$ vibronic states of the ground electronic surface, $X^1\Sigma^+_g$, comprise the quantum register. All ground states are coupled via the dipole coupling to the vibronic levels of the electronically excited surface, $A^1\Sigma^+_u$ (see figure \ref{Na2}).
The final time for the implementation was taken to be $128$ picoseconds.
Figure \ref{Na2_su6_6} summarizes our results for producing an $SU(6)$ Fourier transform, 
$$
FT(6)=\left(\begin{array}{cccccc} 1&1&1& 1&1&1\\
                          1&w&w^2&w^3&w^4&w^5\\
			  1&w^2&w^4&1&w^2&w^4\\
			  1&w^3&1&w^3&1&w^3\\
			  1&w^4&w^2&1&w^4&w^2\\
			  1&w^5&w^4&w^3&w^2&w\\
\end{array}\right);\quad w\equiv\sqrt[6]{1}=e^{2\pi i/6},
$$
on a submanifold of the two potential surfaces containing seven ground state vibrational levels and three excited state vibrational levels. 
The top plot shows the electric field obtained by the local control procedure. The middle plot shows the evolution of the absolute value of the unitary propagator elements under the derived field. The monotonic increase of the objective $\sqrt{J}$ towards its maximal value six can be seen in the bottom plot. The fidelity of the gate, $F=J/N_r^2$,  achieved at the final time, is very close to unity such that $\log_{10}(1-F)\approx-1.7$. Also note that only negligible leakage out of  the quantum register has occurred throughout the process, which is apparent in the flat constant value of the constraint $C$ along the evolution.
A more complete picture of the unitary propagator elements can be obtained from figure \ref{Na2_su6_6_compas}. 
The elements are shown in plots a), b) and c) at times $t=0$, $t=T/2$ and $t=T$ respectively as vectors in the complex plane. At time $t=0$ there are six vectors pointing along the real axis towards unity. These are the diagonal elements of the identity.  The remaining elements are zero. The slight rotation of the  diagonal elements and small population of the offdiagonal elements at initial time are due to the small seed rotation induced on the initial condition to initiate the control algorithm. We stress, however, that due to the smallness of the perturbation, once the field is obtained it can be applied to the desired `pure' initial condition $U(0)=\Bbb{I}$, with negligeble deviations from the current results. As time progresses the elements rotate and shrink/expand to obtain the $SU(6)$ Fourier transform at the final time. It can be clearly seen that the elements at final time (c) divide the circle into six equal parts and thus are, up to an unimportant phase rotation, just powers of $w$, the sixth root of unity appearing in $FT(6)$.

\begin{figure}[ht]
\epsfig{file=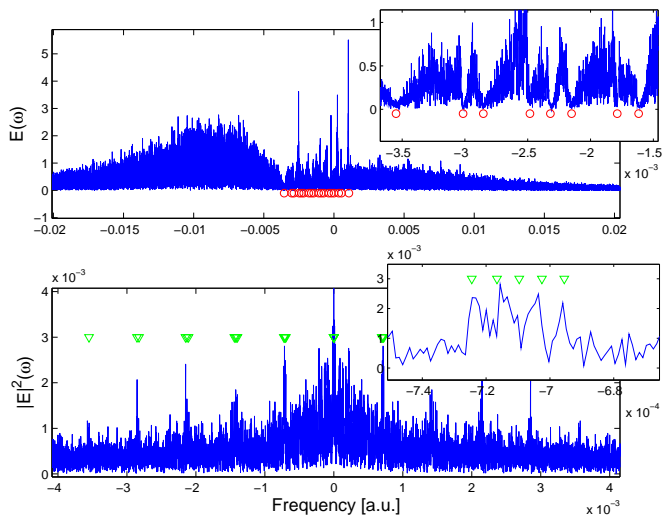,clip=,width=3.5in}
\caption[Na2_su6_6_spect]{\label{Na2_su6_6_spect}   The spectrum $\tilde E(\omega)$  of the field  with one-photon transition frequencies marked by (red) circles (top) and  the spectrum of the field intensity $|\tilde E|^2(\omega)$ with two-photon transition frequencies marked by (green) triangles (bottom).}
\end{figure}

Some understanding of the mechanism by which the leakage is avoided  can be obtained by looking at the spectrum of the field $E(t)$ and its square $|E(t)|^2$ (top and bottom of figure \ref{Na2_su6_6_spect} respectively).
The spectrum $\tilde E(\omega)$ of the field, corresponds to one-photon processes. A glance at the  plot of $\tilde E(\omega)$ (top of  figure \ref{Na2_su6_6_spect}) reveals that there are `holes' in this spectum at precisely the points corresponding to transitions from the quantum register to the excited states,  as indicated by the circles. These `holes' are evidently responsible for the absence of  excitations. As one-photon interactions are suppressed it must be two (and higher) photon processes which produce the evolution towards the desired target. The spectrum $|\tilde E|^2(\omega)$ of the field intensity corresponds to two-photon processes produced by absorption and immediate emission of a photon. The bottom plot of figure  \ref{Na2_su6_6_spect} shows that $|E|^2(\omega)$ displays peaks at the  precise frequencies corresponding to energy differences between the register states, indicated in the figure by green circles. This implies that the field is in two-photon resonance with transitions corresponding to the register states but is detuned from one-photon resonance with excited state transitions.

%%%%%%%%%%%%%%%%%%%%%%%%%%%%%%%%%%%
%\section{Discussion and Conclusions}
The model system used above 
illustrates the general features of our local control method; however it suffers from the following drawbacks. 
%b) $H_0$ motion unaccounted for.
First, the analysis described above was performed in the interaction picture where the $H_0$ motion is transformed away. In practice, however, this drift motion must be taken into account. 
%a)non scalabillity
Second, it is generally expected that the computational power scale exponentially with the physical resources such that for example $N_r$ physical bits carry $2^{N_r}$ entities of information. In the molecular system studied here, as in other studies of molecular quantum computation,  
this requirement is not satisfied since $N_r$ physical levels correspond to only $N_r$ information entities ($\log_2N_r$ bits) namely the scaling is only linear.  
However, the control method we have proposed does not depend on the specific model studied; therefore it can be applied to 
alternative models which are both driftless and scalable. 

In summary, we have shown that it is possible to systematically design fields to produce arbitrary unitary transformations while avoiding leakage from the quantum register. 

We wish to thank Jose Palao and Ronnie Kosloff for helpful discussions.
This work was supported by the US-ONR under grant N00014-01-1-0667 and 
the BMBF-MOS. 

%%%%%%%%%%%%%%%%%%%%%%%%%%%%%%%%
\bibliographystyle{phaip}
\bibliography{shlomo}
%%%%%%%%%%%%%%%%%%%%%%%%%%%%%%

\end{document}